\documentclass[12pt,english,floatfix,nofootinbib,superscriptaddress,aps,prd,preprint]{revtex4}
\usepackage[utf8]{inputenc}
\usepackage{float}
\usepackage{array}
\usepackage{lipsum}
\usepackage{dsfont}
\usepackage{commath}
\usepackage{graphicx}
\usepackage{amsmath,amsthm,amsfonts,amssymb}
\usepackage{graphicx}
\usepackage[english]{babel} 
\usepackage{color}
\usepackage{tensor}
\usepackage{esint}
\usepackage[dvips]{epsfig}
\usepackage[dvips]{graphicx}
\usepackage{float}
\usepackage{units}
\usepackage{textcomp}
\usepackage{mathrsfs}
\usepackage{amsmath}
\usepackage[makeroom]{cancel}
\usepackage{amssymb}
\usepackage{amsbsy}
\usepackage{amsfonts}
\usepackage{amssymb,mathrsfs,xcolor}
\usepackage{esint}
\usepackage{braket}
\usepackage{array}
\usepackage{graphicx}
\usepackage{caption}
\usepackage{subcaption}

\usepackage{wasysym}
\usepackage{multirow}
\usepackage{wrapfig}

\usepackage{stmaryrd}
\usepackage{upgreek}

\makeatletter

\makeatletter\usepackage{babel}

\usepackage{hyperref}
\hypersetup{
    colorlinks,
    citecolor=blue,
    filecolor=green,
    linkcolor=purple,
    urlcolor=red,
}

\usepackage{slashed}

\newcommand{\ie}{\begin{equation}}
\newcommand{\fe}{\end{equation}}
\newcommand{\se}{\begin{eqnarray}}
\newcommand{\ff}{\end{eqnarray}}

\begin{document}

\title{Thermal analysis of photon--like particles in rainbow gravity}


\author{A. A. Ara\'{u}jo Filho}
\email{dilto@fisica.ufc.br}

\affiliation{Departamento de Física Teórica and IFIC, Centro Mixto Universidad de Valencia--CSIC. Universidad
de Valencia, Burjassot-46100, Valencia, Spain}

\author{J. Furtado}
\email{job.furtado@ufca.edu.br}

\affiliation{Universidade Federal do Cariri, Centro de Ciências e Tecnologia, 63048-080, Juazeiro do Norte, CE, Brazil}

\author{H. Hassanabadi}
\email{hha1349@gmail.com}

\affiliation{Department   of   Physics,   University   of   Hradec   Kr\'{a}lov\'{e}, Rokitansk\'{e}ho   62,   500   03   Hradec   Kr\'{a}lov\'{e},   Czechia}

\affiliation{Physics Department, Shahrood University of Technology, Shahrood, Iran}

\author{J. A. A. S. Reis}
\email{jalfieres@gmail.com}

\affiliation{Universidade Estadual do Sudoeste da Bahia (UESB), Departamento de Ciências Exatas e Naturais, Campus Juvino Oliveira, Itapetinga -- BA, 45700-00,--Brazil}


\date{\today}

\begin{abstract}

This work is devoted to study the thermodynamic behavior of photon--like particles within the \textit{rainbow} gravity formalism. To to do this, we chose two particular ansatzs to accomplish our calculations. First, we consider a dispersion relation which avoids UV divergences, getting a positive effective cosmological constant. We provide \textit{numerical} analysis for the thermodynamic functions of the system and bounds are estimated. Furthermore, a phase transition is also expected for this model.
Second, we consider a dispersion relation employed in the context of \textit{Gamma Ray Bursts}. Remarkably, for this latter case, the thermodynamic properties are calculated in an \textit{analytical} manner and they turn out to depend on the harmonic series $H_{n}$, gamma $\Gamma(z)$, polygamma $\psi_{n}(z)$ and zeta Riemann functions $\zeta(z)$.


\end{abstract}

\maketitle


\section{Introduction}

\textit{Rainbow} gravity is a promising approach for creating self--consistent cosmological models that evade singularities \cite{majumder2013singularity}, which lack clear definitions within the framework of quantum gravity \cite{kangal2022effective,kangal2021relativistic,r1,r2,r3,r4,r5,r6,r8,r9}. The choice of \textit{rainbow} functions can lead to varying predictions regarding the evolution of the universe, including acceleration of cosmic expansion \cite{sefiedgar2017entropic,brighenti2017primordial,nojiri2007introduction,amelino2013rainbow,sefiedgar2018thermodynamics}. In this manner, exploring a diverse range of them is essential for both fundamental physics and cosmology \cite{mota2019combined,younesizadeh2021new,fomin2020exact}. Through this investigation, we can gain new perspectives on the behavior of the universe and deepen our understanding of the intricate interplay between quantum gravity and cosmology, potentially leading to novel results and approaches for understanding the fundamental nature of our universe.

The study of thermodynamics in the context of quantum gravity is essential for comprehending the behavior of physical systems under the effects of modifications to the dispersion relation, including those described by \textit{rainbow} gravity \cite{dehghani2018thermal,dehghani2019thermodynamic,sefiedgar2017thermodynamics,haldar2019thermodynamics,hamil2022effect,kim2016thermodynamic,dehghani2018thermodynamics,araujo2021bouncing,sefiedgar2018thermodynamics}. The non--trivial thermodynamic properties resulting from these modifications can include changes to the thermodynamic equation of state, the critical temperature, the emergence of novel phase transitions and modifications to the Bose--Einstein condensation \cite{feng2017thermodynamic,dehghani2020ads4,md2018phase, Furtado:2021aod}.

In addition, this exploration is of significant interest, particularly for its implications for black hole behavior. One of the key consequences of the modification of the dispersion relation in this framework is the emergence of a minimum length scale \cite{nilsson2017energy,roy2023entropy}, which leads to modifications of both the Hawking temperature and the entropy of black holes \cite{feng2020rainbow,yekta2019joule}. As a result, new proposals have arisen for black hole thermodynamics, including modified entropy--area relations and first laws of thermodynamics \cite{feng2018rainbow,sefiedgar2017entropic,sefiedgar2016can}. These novel ideas offer a fresh perspective on the behavior of black holes and could potentially pave the way for a deeper understanding of them.

Here, we study the thermodynamic behavior of photon--like particles within the \textit{rainbow} gravity formalism. To to do this, we chose two particular ansatzs to accomplish our calculations. First, we consider a dispersion relation which avoids UV divergences, getting a positive effective cosmological constant. To this case, we provide \textit{numerical} analysis for the thermodynamic functions of the system. Also, new bounds are estimated and a phase transition is expected to our model, being highlighted by the heat capacity behavior. Second, we consider a dispersion relation employed in the context of \textit{Gamma Ray Bursts}, which present a remarkable feature: \textit{analytical} results. The thermal quantities turn out to depend on the following special functions: harmonic series $H_{n}$, gamma $\Gamma(z)$, polygamma $\psi_{n}(z)$ and zeta Riemann functions $\zeta(z)$.

This paper is organized as follows: in Sec. \ref{sec2}, we briefly present the modified \textit{rainbow} dispersion relation. In section \ref{sec3}, we investigate the thermodynamic properties for two \textit{rainbow} functions and new bounds are estimated. Finally, in section \ref{conclusion}, we outline our conclusion.


\section{The modified dispersion relation} \label{sec2}

The general dispersion relation of massive particles in the context of \textit{Double special relativity} reads \cite{amelino2001testable,amelino2002relativity}
\ie
E^{2}g_{1}^{2}(E/E_{P})-k^{2}g_{2}^{2}(E/E_{P}) = m^{2},\label{maindispersionrelation}
\fe
where $g_{1}(E/E_{P})$ and $g_{2}(E/E_{P})$ are simply arbitrary functions whose the following conditions must be satisfied
\ie
\lim_{E/E_{P} \rightarrow 0}g_{1}(E/E_{P}) = 1, \,\,\,\,\,\, \text{and} \,\,\,\,\,\,\lim_{E/E_{P} \rightarrow 0}g_{2}(E/E_{P}) = 1.
\fe
Notice that the usual dispersion relation is recovered when low energy regime is considered.
The first ideas of \textit{Double special relativity} were developed to flat spacetimes. Nevertheless, it is not prohibited to consider a more general curved background, accounting for general relativity. Within this viewpoint, Magueijo and Smolin \cite{magueijo2004gravity} suggested that the Einstein’s field equations as well as the stress--energy tensor should be modified by introducing a new parameter to the following equations
\ie
G_{\mu\nu}(E) = 8 \pi G(E) T_{\mu\nu}(E) + g_{\mu\nu}\Lambda(E),
\fe
where $G_{\mu\nu}$ is the Einstein tensor and
$G_{\mu\nu} = R_{\mu\nu}-\frac{1}{2} R g_{\mu\nu}$, being $R_{\mu\nu}$ the Ricci curvature tensor and $R$ the scalar curvature; $\Lambda(E)$ and $G(E)$ are the energy--dependent cosmological and Newtons constant respectively. $G(0)$ can also be defined as the usual low--energy Newton constant. In a similar manner, we may also have the line element of the \textit{rainbow}--like Schwarzschild black hole
\ie
\mathrm{d}s^{2} = - \left( 1 - \frac{2MG(0)}{r}\right)\frac{\mathrm{d}t^{2}}{g_{1}^{2}(E/E_{P})} + \frac{\mathrm{d}r^{2}}{(1-2MG(0)/r)g_{2}^{2}(E/E_{P})} + \frac{r^{2}(\mathrm{d}\theta^{2}+\sin^{2}\theta \mathrm{d}\phi^{2})}{g_{2}^{2}(E/E_{P})}.\label{metric}
\fe

Since the effects of $g_{1}(E/E_{P})$ and $g_{2}(E/E_{P})$ are highlighted when the energy $E$ is similar to $E_{P}$, they modify the ultraviolet behavior analogous to what happens to the
non--commutative geometry scenarios and the generalized uncertainty principle. If its modification plays a role in the Liouville measure $\mathrm{d}^{3}x\mathrm{d}^{3}k$, Eq. (\ref{metric}) may naturally gives rise to ultraviolet regulator. This feature is brought about due to the presence of appropriate choices of functions $g_{1}(E/E_{P})$ and $g_{2}(E/E_{P})$. 
Corroborating these arguments, a notable results have been obtained, applying, \textit{rainbow} gravity to the black hole entropy \cite{garattini2010modified}. In this reference, the UV regulator, called ``the brick wall'', has been removed by considering the following choice of $g_{1}(E/E_{P})$ and $g_{2}(E/E_{P})$: $g_{1}(E/E_{P})/g_{2}(E/E_{P})= \exp[-E/E_{P}]$. Nevertheless, as it has been argued in Ref. \cite{garattini2011modified}, this particular choice turns out to give us a negative effective cosmological constant. In order to overcome this situation, we shall provide our investigations based on another dispersion relation that ``cure'' such an issue. This and other feats will be discussed with more details in the next section.


\section{Thermodynamic properties} \label{sec3}

\subsection{The first case}

As pointed out in Refs. \cite{garattini2005casimir,garattini2006cosmological}, there exists an appearance of an induced cosmological constant $\Lambda/8\pi G$. However, it does not address neither a renormalization nor a regularization for the sake of avoiding UV divergences. Notice that if one takes into account the pure Gaussian regulator only, one verifies that the \textit{zero point energy} is fundamentally negative for \textit{rainbow} gravity \cite{garattini2011modified}. Nevertheless, after some parametrizations, Ref. \cite{garattini2011modified} got a positive value to the effective cosmological constant among other features by using the following ansatz:
\ie
 g_{2}(E/E_{P})=1,\phantom{a} g_{1}(E/E_{P})  = \sum^{n}_{i=0} \xi_{i}\left(\frac{E}{E_{P}}\right)^{i}\exp\left(-\alpha\frac{E^{2}}{E^{2}_{P}}\right) \phantom{a} \forall \phantom{a} \alpha>0, \xi_{i} \in \mathbb{R}. \label{gs}
\fe
It is exactly with above expression that we shall focus on in order to provide our calculations. Next, combining Eq. (\ref{maindispersionrelation}) with Eq. (\ref{gs}), we obtain
\ie
E^{2} \left[ \sum^{n}_{i=0} \xi_{i}\left(\frac{E}{E_{P}}\right)^{i}\exp\left(-\alpha \frac{E^{2}}{E^{2}_{P}}\right)\right]^{2} - k^{2} = m^{2}.
\fe
Although we have presented a general modified dispersion relation, to perform our calculation, we shall particularize it, namely, $m=0$ (massless particles) and $i=0$. In other words, we shall focus on the simplest massless case. Such a case leads to the following partition function
\ie
\ln[Z] = -\frac{1}{2} \ln \left(1-e^{-\beta  E}\right) \left(2 E \xi _0^2 e^{-\frac{2 \alpha  E^2}{E_P^2}}-\frac{4 \alpha  E^3 \xi _0^2 e^{-\frac{2 \alpha  E^2}{E_P^2}}}{E_P^2}\right) \sqrt{E^2 \xi _0^2 e^{-\frac{2 \alpha  E^2}{E_P^2}}},
\fe
where $\beta \equiv 1/\kappa_{B} T$, and $\kappa_{B}$ is the Boltzmann constant. After that, all thermodynamic quantities can be addressed. It is important to note that, from now on, the thermodynamic properties will be calculated, taking into account the following constant values $\xi_{0}=\alpha=1$.


\subsubsection{Equation of states}

\begin{figure}
    \centering
    \includegraphics[scale=0.4]{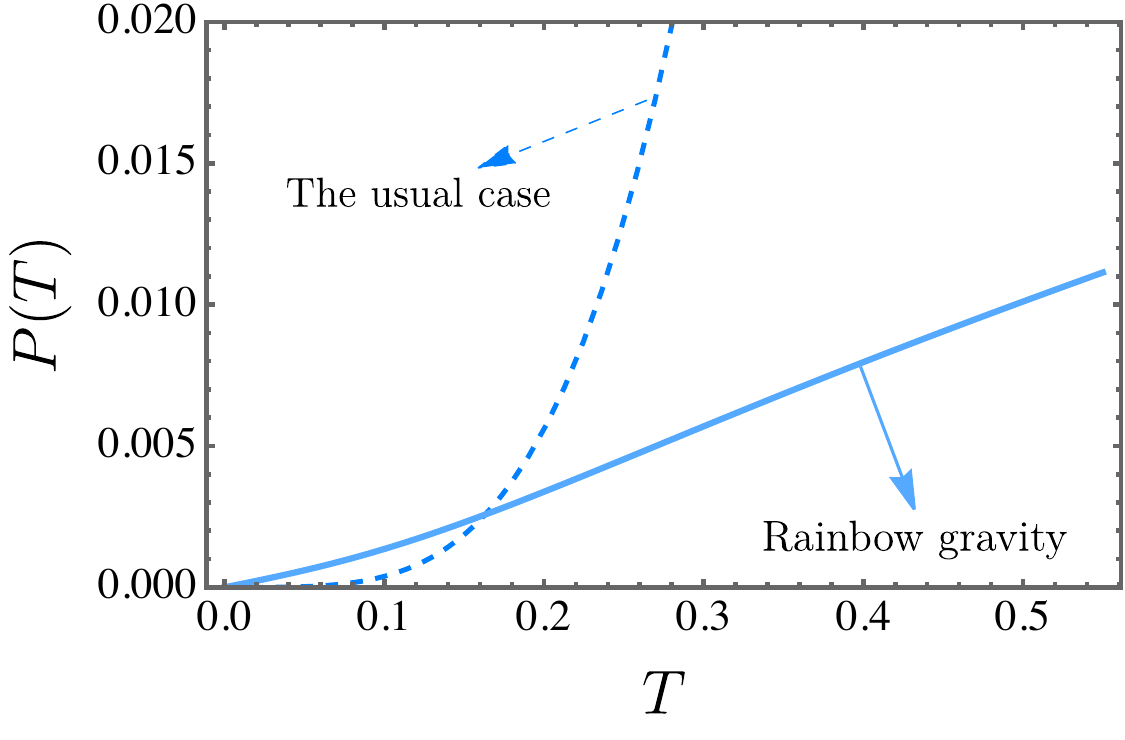}
    \includegraphics[scale=0.4]{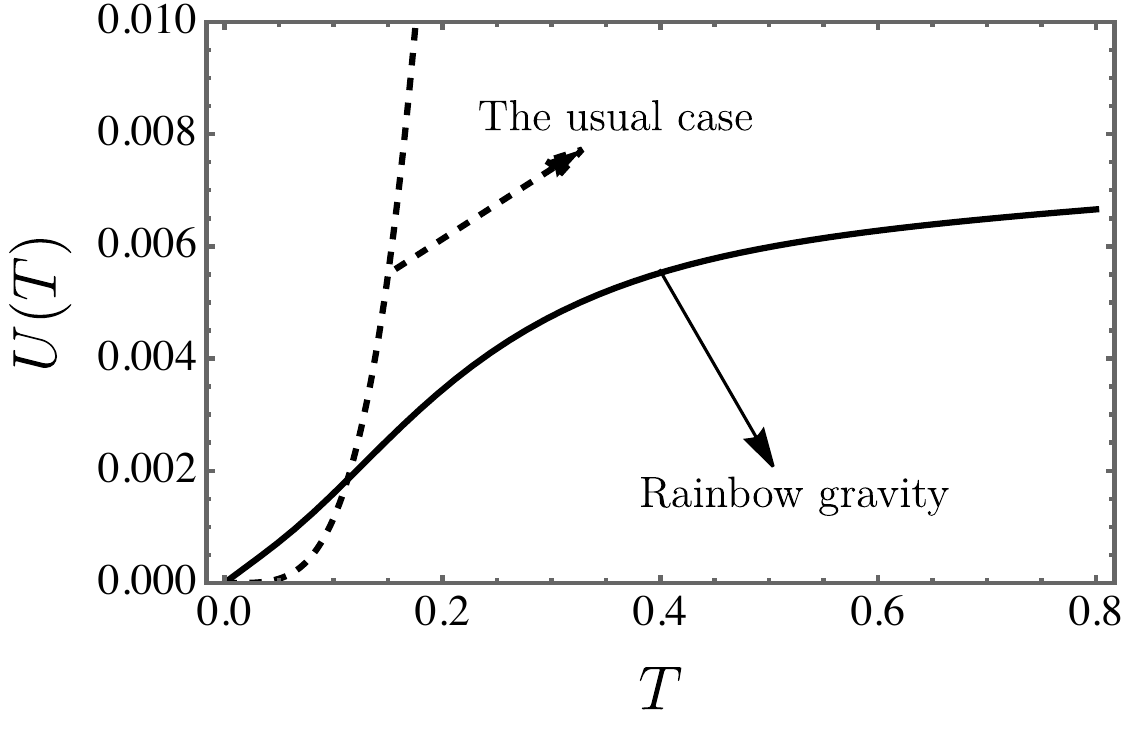}
    \includegraphics[scale=0.4]{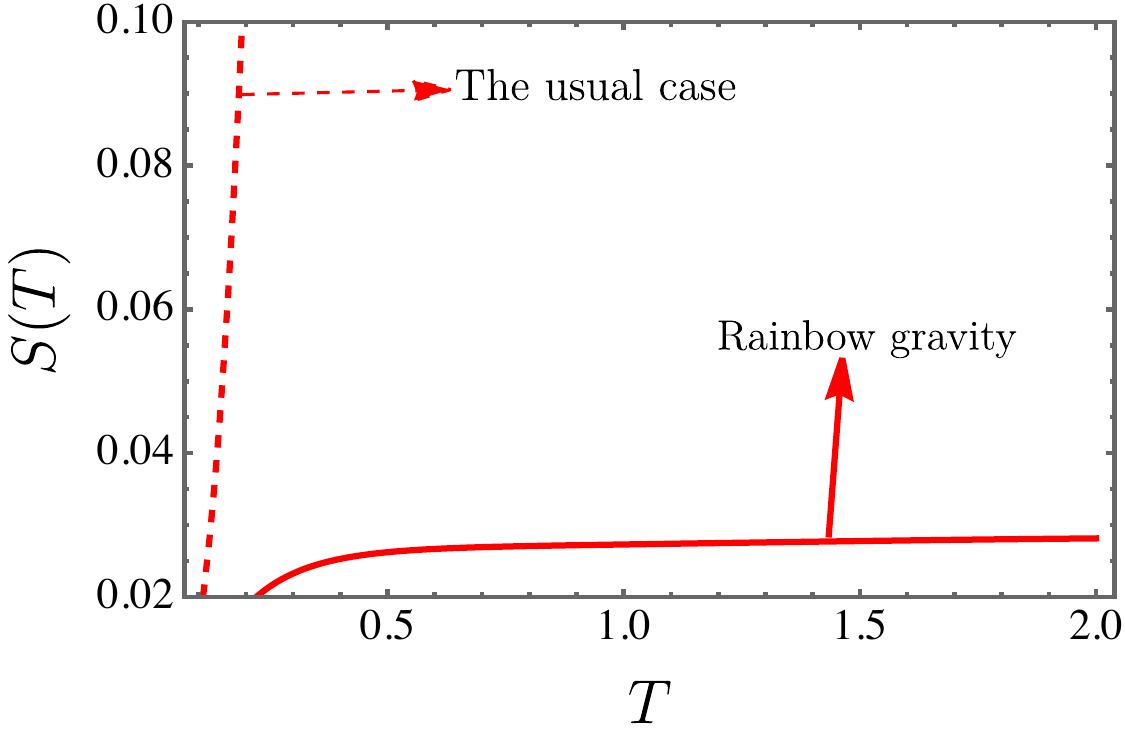}
    \includegraphics[scale=0.4]{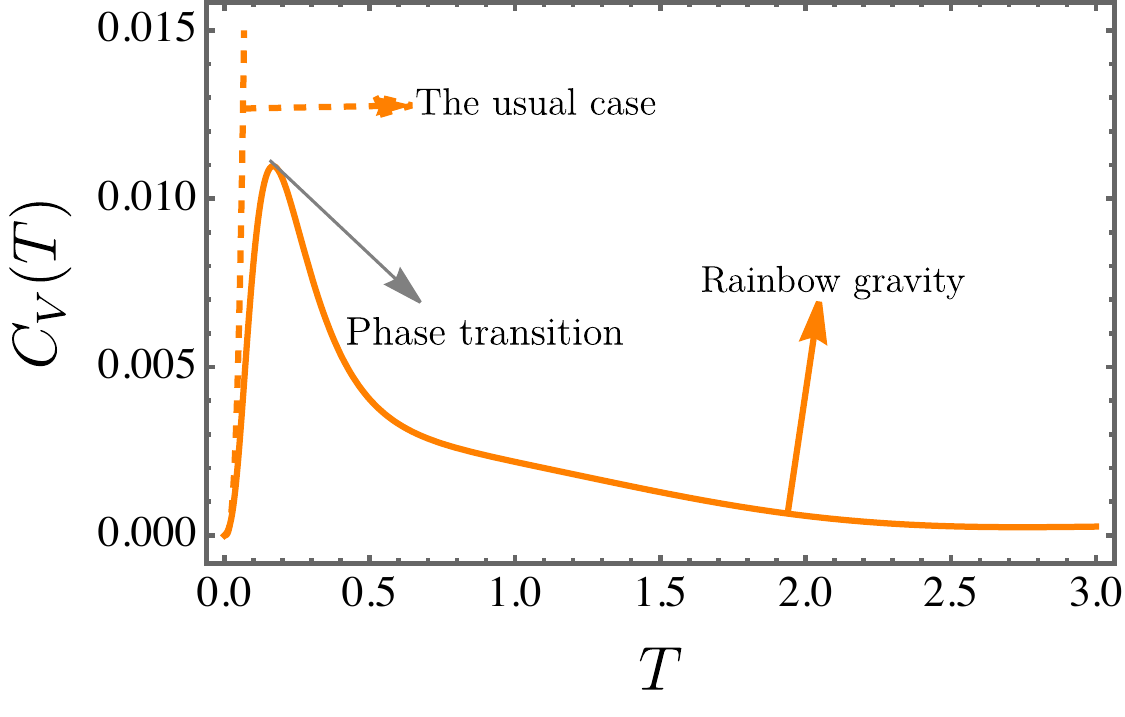}
    \caption{All thermodynamic state quantities for different values of temperature $T$. The point that the pressures coincide is $T=0.11$ GeV; also, the point the mean energies coincide is $T=0.16$ GeV.}
    \label{all}
\end{figure}

In this subsection, we focus on the study of the equation of states. Then, it reads
\ie
P = -\int^{\infty}_{0}\frac{\ln \left(1-e^{-\beta  E}\right) }{2 \beta}\sqrt{\xi_0^2 E^2 e^{-\frac{2 \alpha  E^2}{E_P^2}}} \left(2 \xi_0^2 E e^{-\frac{2 \alpha  E^2}{E_P^2}}-\frac{4 \alpha  \xi_0^2 E^3 e^{-\frac{2 \alpha  E^2}{b^2}}}{E_P^2}\right) \mathrm{d}E.\label{Pressure}
\fe
Here, we notice that we have $lim_{\alpha \rightarrow \infty}P = 0$, and
\ie
\lim_{\alpha \rightarrow 0}P = \frac{\pi ^4}{45} \left(\xi_0^2\right){}^{3/2}T^{4} = \frac{\pi^{2}}{3}\sigma \left(\xi_0^2\right){}^{3/2}T^{4} = \Tilde{\alpha}T^{4}, \label{mdr1}
\fe
where $\Tilde{\alpha} = \pi^{2}\sigma \left(\xi_0^2\right){}^{3/2}/3$ and $\sigma$ is the well--known \textit{Stefan--Boltzmann} constant given by
\ie
\sigma = \frac{1}{\pi^{2}}  \int^{\infty}_{0}  \frac{E^{3}\,e^{-\beta E}} {\left(  1- e^{-\beta E} \right)} \mathrm{d}E = \frac{\pi^{2}}{15}.
\label{radiance}
\fe
It is important to highlight, even considering such limits to the massive case, the integral did not converge. This is why we opted to work on the massless particles instead in this manuscript. Also, Eq. (\ref{mdr1}) gives rise to a modified \textit{Stefan--Boltzmann} law. As we may probably realize, Eq. (\ref{Pressure}) does not possess \textit{analytical} results. To overcome this issue, we provide a \textit{numerical} analysis as seen in Fig. \ref{all}. This thermodynamic property is a monotonically increasing function for different values of temperature $T$. Also, Fig. \ref{all} provides a clear visualization of the impact of \textit{rainbow} gravity in pressure as temperature increase compared to the conventional case where the effects of \textit{rainbow} gravity are disregarded. Notably, the disparity between the two cases is more pronounced at higher temperatures, which is expected as the\textit{rainbow} gravity influence becomes increasingly significant at elevated energy levels.

\subsubsection{Mean energy}

\begin{figure}
    \centering
    \includegraphics[scale=0.4]{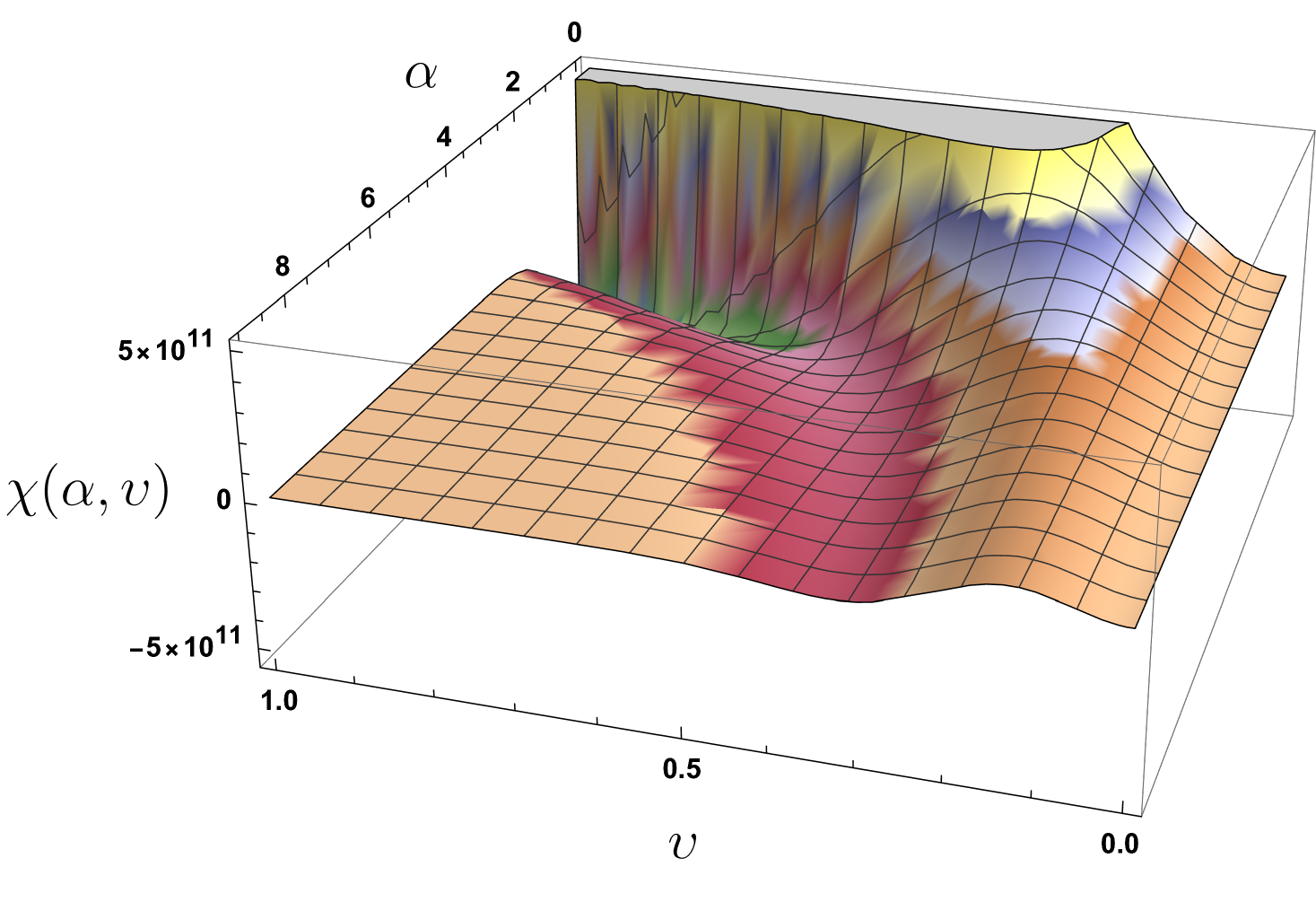}
    \caption{The black body radiation for different values of $\alpha$ and $\upsilon$.}
    \label{radiation}
\end{figure}

Here, we investigate the main aspects of the mean energy for our system. In addition, the black body radiation can also be examined as well in possession of this thermal state quantity. With this purpose, we write
\ie
U = \int^{\infty}_{0} \frac{E e^{-\beta  E}}{2 \left(1-e^{-\beta  E}\right)} \left(2 E \xi _0^2 e^{-\frac{2 \alpha  E^2}{E_P^2}}-\frac{4 \alpha  E^3 \xi _0^2 e^{-\frac{2 \alpha  E^2}{E_P^2}}}{E_P^2}\right) \sqrt{E^2 \xi _0^2 e^{-\frac{2 \alpha  E^2}{E_P^2}}}  \mathrm{d}E.
\fe
It is worthy to be mentioned that $lim_{\alpha \rightarrow \infty}U = 0$, and $lim_{\alpha \rightarrow 0}U$ results
\ie
\lim_{\alpha \rightarrow 0}U = \frac{\pi^4}{15} \left(\xi_0^2\right){}^{3/2}T^{4} = \pi^{2}\sigma \left(\xi_0^2\right){}^{3/2}T^{4} = \Bar{\alpha}T^{4},\label{meanenergyeq}
\fe
where $\Bar{\alpha}\equiv \pi^{2}\sigma\left(\xi_0^2\right){}^{3/2}$. Similarly with what happened to the equation of states, here, we have recovered the \textit{Stefan--Boltzmann} law after redefining the constants. For a more robust analysis besides this one given by this particular limit, we show a \textit{numerical} analysis to the mean energy. The results are shown in Fig. \ref{all}. As we can see from the plot, the mean energy increases its values when the temperature varies until tending to reach a constant behavior at $T>0.8$ GeV, which contrast to the usual mean energy behavior, i.e., monotonically increasing function. More so, in a straightforward manner, from Eq. (\ref{meanenergyeq}), we can also obtain the associated black body radiation
\ie
\chi(\alpha,\upsilon) = \frac{h\nu e^{-\beta  h\nu}}{2 \left(1-e^{-\beta  h\nu}\right)} \left(2 h\nu \xi _0^2 e^{-\frac{2 \alpha  h\nu^2}{E_P^2}}-\frac{4 \alpha  h\nu^3 \xi _0^2 e^{-\frac{2 \alpha  h\nu^2}{E_P^2}}}{E_P^2}\right) \sqrt{h\nu^2 \xi _0^2 e^{-\frac{2 \alpha  h\nu^2}{E_P^2}}},
\fe
where $h$ is the Planck constant and $\nu$ is the frequency. The corresponding behavior of such a quantity is displayed in Fig. \ref{radiation}. It is worth highlighting that one of the parameters which account for the \textit{rainbow} gravity, namely, $\alpha$, plays an important role in the shape of the black body radiation. Furthermore, if the following limit is considered $\lim_{\alpha \rightarrow 0}X$, we obtain
\ie
\lim_{\alpha \rightarrow 0}X = \frac{(h\nu)^2 \xi _0^2 e^{-\beta  h\nu} \sqrt{(h\nu)^2 \xi _0^2}}{1-e^{-\beta  h\nu}}.
\fe
Notice that, if parameter $\xi_{0}\rightarrow 1$, the usual well--known result encountered in the literature for the black body radiation is recovered.

Now let us estimate a bound for the $\xi_{0}$ parameter comparing to some black body radiation spectrum. Let us consider that $\xi_{0}\approx 1+\delta\xi_{0}$, so that
\ie
U \approx \pi^{2}\sigma \left[(1+\delta\xi_{0})^{2}\right]{}^{3/2}T^{4} \approx \pi^{2}\sigma T^{4}+3\pi^{2}\sigma\delta\xi_{0} T^{4}.\label{meanenergyeq11}
\fe
Based on the above expansion, let us now try to find a bound to $\delta\xi_{0}$. To
do so, we suppose the contribution of $\delta\xi_{0}$ is less than the observational error. From Ref. \cite{BBRadiations}, we can make such an estimation via the microwave background radiation using black body radiation inversion:
\ie
\delta\xi_{0}\leq 1.50\times 10^{-12}. \label{meanenergyeq12}
\fe


\subsubsection{Entropy}

The entropy in this section will be discussed. It reads
\ie
\begin{split}
S = &\int^{\infty}_{0} \beta^{2} \left[\frac{E e^{-\beta  E} \sqrt{E^2 \xi _0^2 +e^{-\frac{2 \alpha  E^2}{E_P^2}}} \left(2 E \xi _0^2 e^{-\frac{2 \alpha  E^2}{E_P^2}}-\frac{4 \alpha  E^3 \xi _0^2 e^{-\frac{2 \alpha  E^2}{E_P^2}}}{E_P^2}\right)}{2 \beta  \left(1-e^{-\beta  E}\right)} \right.\\
& \left. -\frac{\ln \left(1-e^{-\beta  E}\right) \sqrt{E^2 \xi _0^2 +e^{-\frac{2 \alpha  E^2}{E_P^2}}} \left(2 E \xi _0^2 e^{-\frac{2 \alpha  E^2}{E_P^2}}-\frac{4 \alpha  E^3 \xi _0^2 e^{-\frac{2 \alpha  E^2}{E_P^2}}}{E_P^2}\right)}{2 \beta ^2}\right]\mathrm{d}E.
\end{split}
\fe
Analogously what we have done with the previous thermodynamic functions, we take into account $lim_{\alpha \rightarrow \infty}S = 0$, and $lim_{\alpha \rightarrow 0}S$ leads to
\ie
S = \frac{4 \pi ^4}{45} \left(\xi _0^2\right){}^{3/2}T^{3}.
\fe
In order to obtain a general panorama, we provide \textit{numerical} analysis to this thermodynamic state function as well, whose the plot is shown in Fig. \ref{all}. Here, we clearly see that this thermodynamic quantity almost reaches a constant behavior in comparison with the usual case, which is a monotonically increasing function to different values of temperature $T$.

\subsubsection{Heat capacity}

Finally, to complete our investigation, we supply the last thermal function, the heat capacity. In this sense, we can write it as
\ie
\begin{split}
C_{V} =  & \int^{\infty}_{0} \beta^{2} \left[  \frac{E^2 e^{-2 \beta  E} \sqrt{E^2 \xi _0^2 +e^{-\frac{2 \alpha  E^2}{E_P^2}}} \left(2 E \xi _0^2 e^{-\frac{2 \alpha  E^2}{E_P^2}}-\frac{4 \alpha  E^3 \xi _0^2 e^{-\frac{2 \alpha  E^2}{E_P^2}}}{E_P^2}\right)}{2 \left(1-e^{-\beta  E}\right)^2}    \right. \\
& \left.  + \frac{E^2 e^{-\beta  E} \sqrt{E^2 \xi _0^2 +e^{-\frac{2 \alpha  E^2}{E_P^2}}} \left(2 E \xi _0^2 e^{-\frac{2 \alpha  E^2}{E_P^2}}-\frac{4 \alpha  E^3 \xi _0^2 e^{-\frac{2 \alpha  E^2}{E_P^2}}}{E_P^2}\right)}{2 \left(1-e^{-\beta  E}\right)} \right] \mathrm{d}E.
\end{split}
\fe
Next, we consider $lim_{\alpha \rightarrow \infty}C_{V} = 0$, and $lim_{\alpha \rightarrow 0}C_{V}$ leads to
\ie
C_{V} = \frac{4 \pi ^4}{15} \left(\xi _0^2\right){}^{3/2}T^{3}.
\fe
Now, we supply a \textit{numerical} analysis to heat capacity, which is exhibited in Fig. \ref{all}. In comparison with the previous thermal quantity, an analogous behavior also occurs to the usal case: the existence of an monotonically increasing function when the temperature $T$ runs. Nevertheless, to the \textit{rainbow} case we verify that actually does exist a phase transition for $T=0.167$ GeV, i.e., at the point $(0.167,0.010)$; and the second law of thermodynamics is still maintained \cite{araujo2023thermodynamics,furtado2023thermodynamical}. It worth commenting that the similar studies have also appeared in the literature recently \cite{aa1,aa2,aa3,aa4,aa5,aa6,aa7,aa8,aa9,aa10,aa11,aa12,aa13,aa14}.


\subsection{The second case}

The selection of next \textit{rainbow} functions was initially explored in \cite{51} within the framework of \textit{Gamma Ray Bursts}. Later, this particular choice of \textit{rainbow} functions was further studied with respect to its application to FRW solutions \cite{52,53}.

Here, we present the subsequent ansazts that we shall use to perform our calculation
\ie
g_{1}(E/E_{P}) =\frac{e^{\frac{E \xi }{E_{P}}}-1}{\frac{E \xi }{E_{P}}}, \,\,\,\,\,\, g_{2}(E/E_{P}) = 1  \label{main2}. 
\fe
Similarly to what we have done in the previous section, now we shall investigate the thermodynamic behavior of the latter case present in Eq. (\ref{main2}). To this case, the partition function reads
\ie
\ln[Z_{2}] = -\int^{\infty}_{0}\frac{E_{P}^2}{\xi ^2} e^{\frac{E \xi }{E_{P}}} \left(e^{\frac{E \xi }{E_{P}}}-1\right)^2 \ln \left(1-e^{-\beta  E}\right) \mathrm{d}E.
\fe

\subsubsection{Equation of states}

\begin{figure}
    \centering
    \includegraphics[scale=0.4]{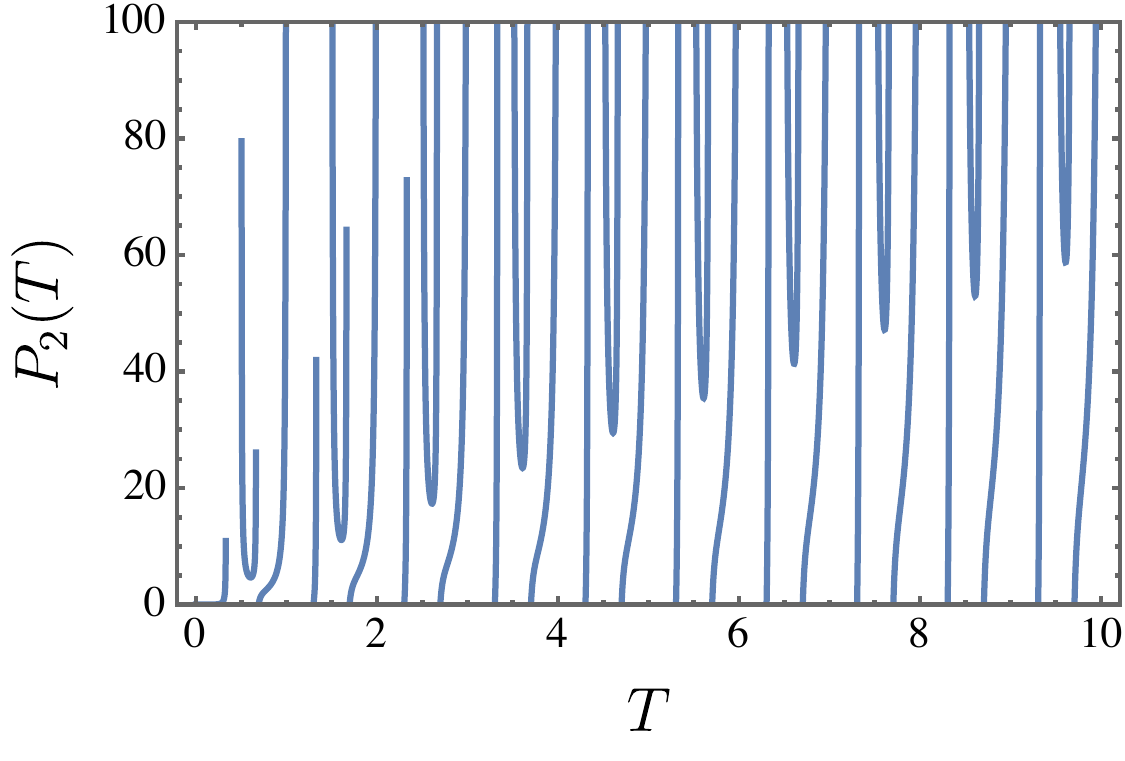}
    \includegraphics[scale=0.4]{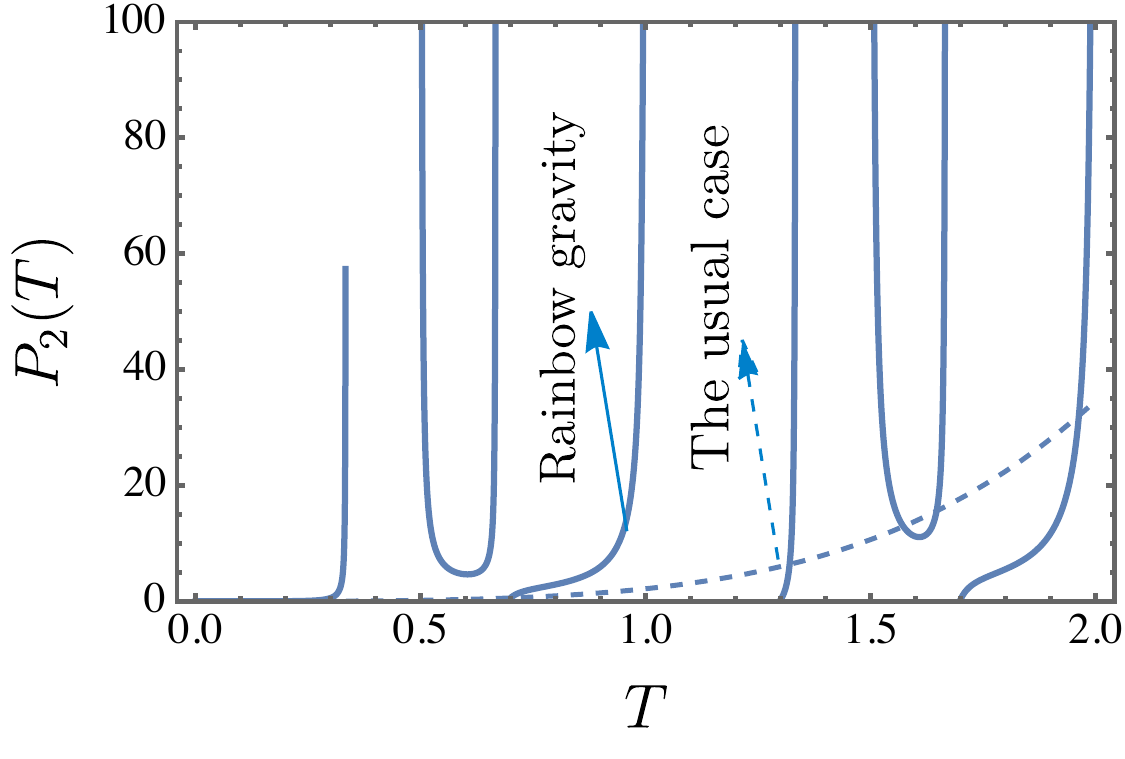}
    \caption{Pressure for different values of temperature $T$, considering the second case.}
    \label{Pressure2}
\end{figure}

The study of the equation of states in the context of \textit{rainbow} gravity is motivated by the need to understand how gravity behaves under the influence of quantum effects. They characterize the behavior of matter and energy in a gravitational system. By studying it in the context of \textit{rainbow} gravity, we aim to gain insights into the nature of gravity at the quantum level. With this, we can possibly address new phenomenology for describing the behavior of black holes and neutron stars for instance. In this sense, the equation of states are straightforwardly derived as
\ie
\begin{split}
P_{2}(T) &= -\int^{\infty}_{0} \frac{E_{P}}{\beta  \xi ^2} e^{\frac{E \xi }{E_{P}}} \left(e^{\frac{E \xi }{E_{P}}}-1\right) \sqrt{a^2 \left(e^{\frac{E \xi }{E_{P}}}-1\right)^2} \ln \left(1-e^{-\beta  E}\right)  \mathrm{d}E \\
 &= -\frac{E_{P}^{3} \left(3 H_{-\frac{\xi }{E_{P} \beta }}-3 H_{-\frac{2 \xi }{E_{P} \beta }}+H_{-\frac{3 \xi }{E_{P} \beta }}\right)}{3 \beta  \xi ^3},
 \end{split}
\fe
where $H_{n}$ is the harmonic series defined by
\ie
H_{n} = \sum_{n}^{k=1} \frac{1}{k}.
\fe
In Fig. \ref{Pressure2}, we display the behavior of $P_{2}(T)$. Due to the harmonic series characteristic, the pressure shows discontinuity in its shape. The presence of such a discontinuity might be seen as a signal that the current assumption or ansatz for the \textit{rainbow} functions is not physically valid within this context. As we shall see, a similar periodic discontinuity is also evident in other thermodynamic quantities.


\begin{figure}
    \centering
    \includegraphics[scale=0.4]{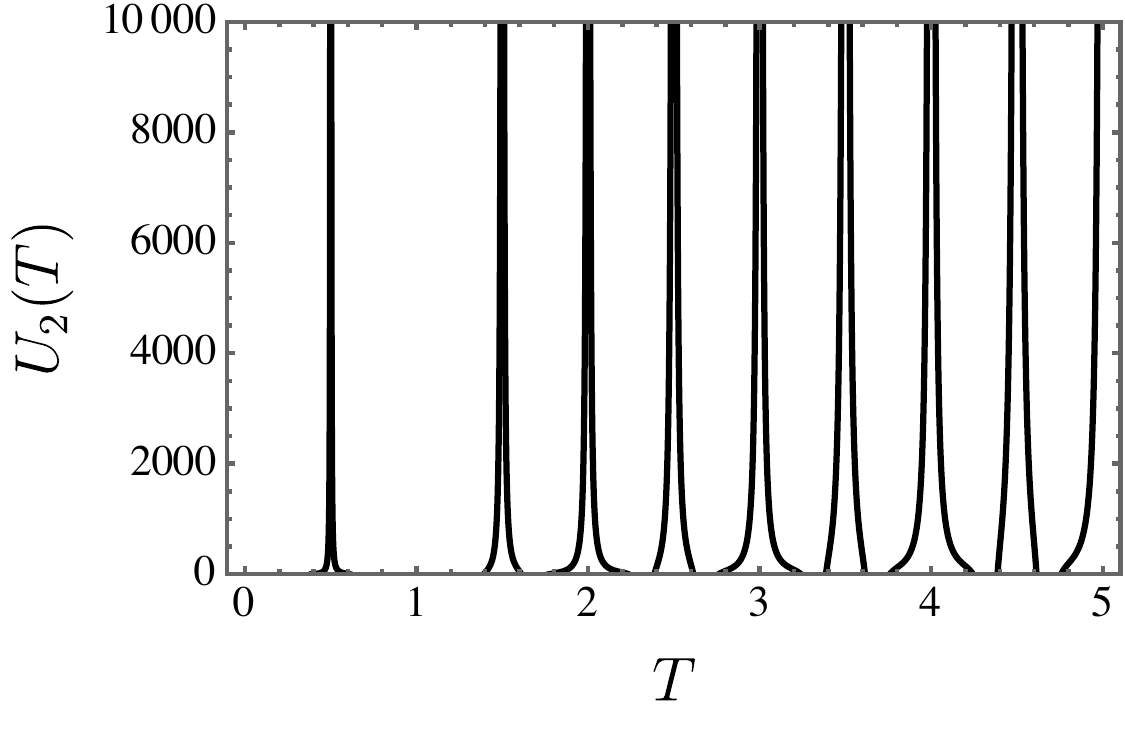}
    \includegraphics[scale=0.4]{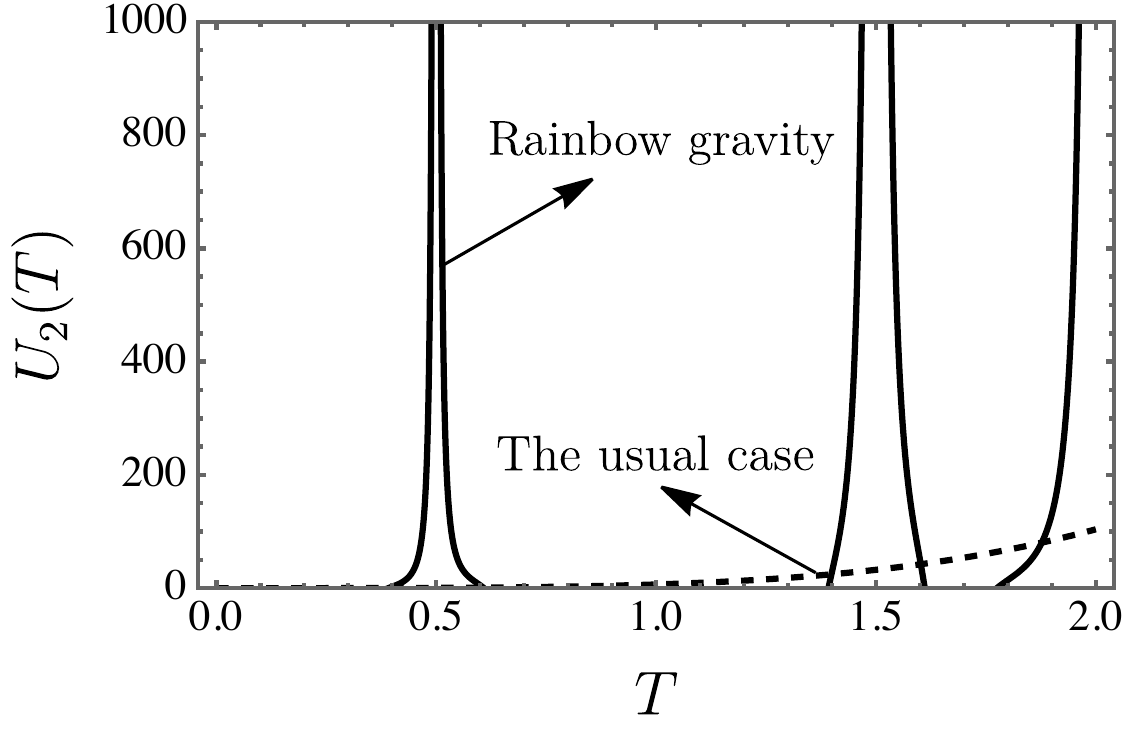}
    \caption{Mean energy for different values of temperature $T$, considering the second case.}
    \label{Energy2}
\end{figure}

\begin{figure}
    \centering
    \includegraphics[scale=0.4]{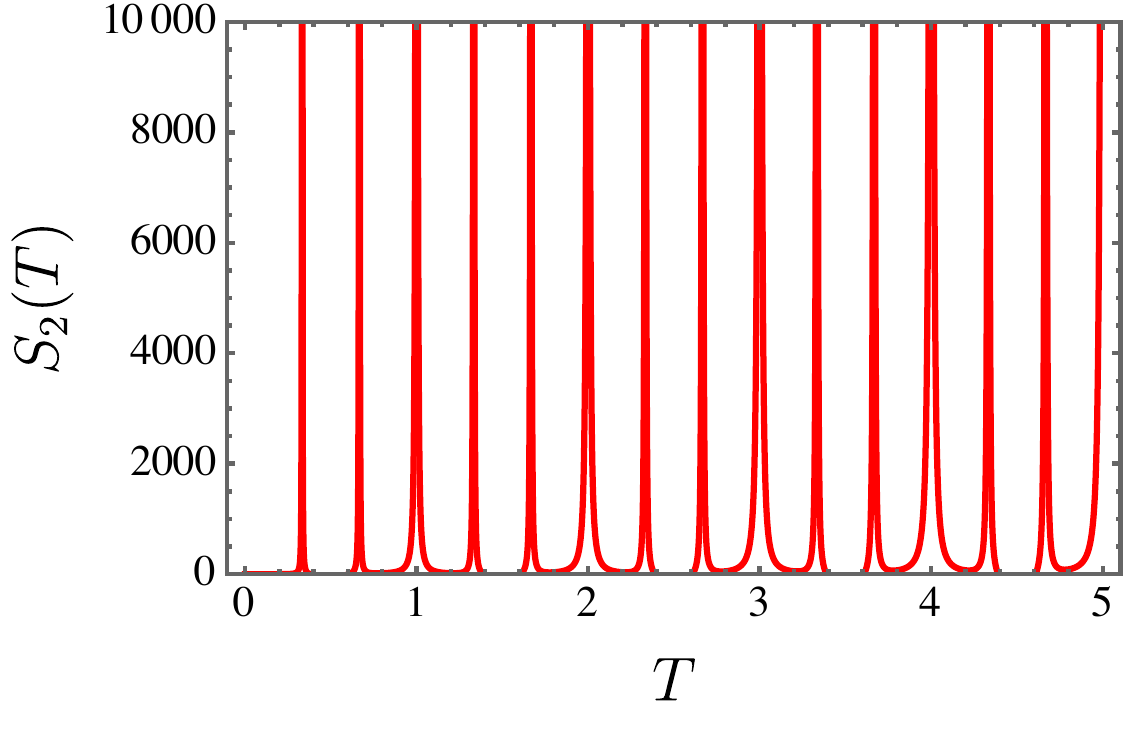}
    \includegraphics[scale=0.4]{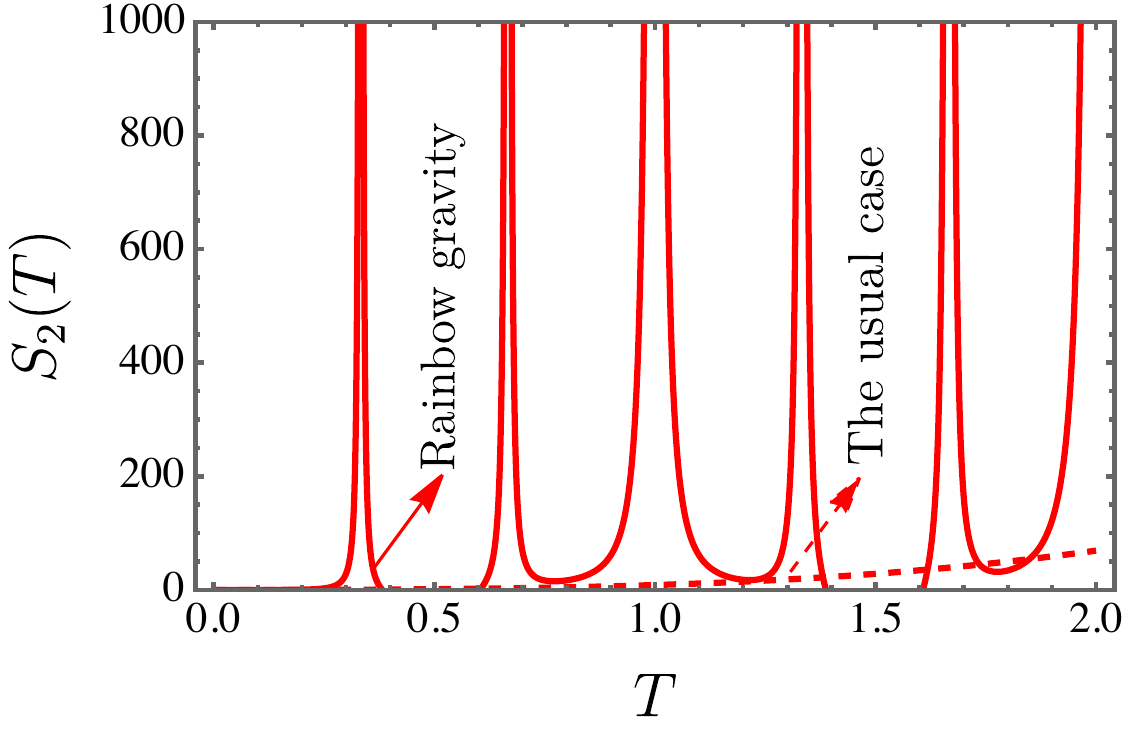}
    \caption{Entropy for different values of temperature $T$, considering the second case.}
    \label{Entropy2}
\end{figure}

\begin{figure}
    \centering
    \includegraphics[scale=0.41]{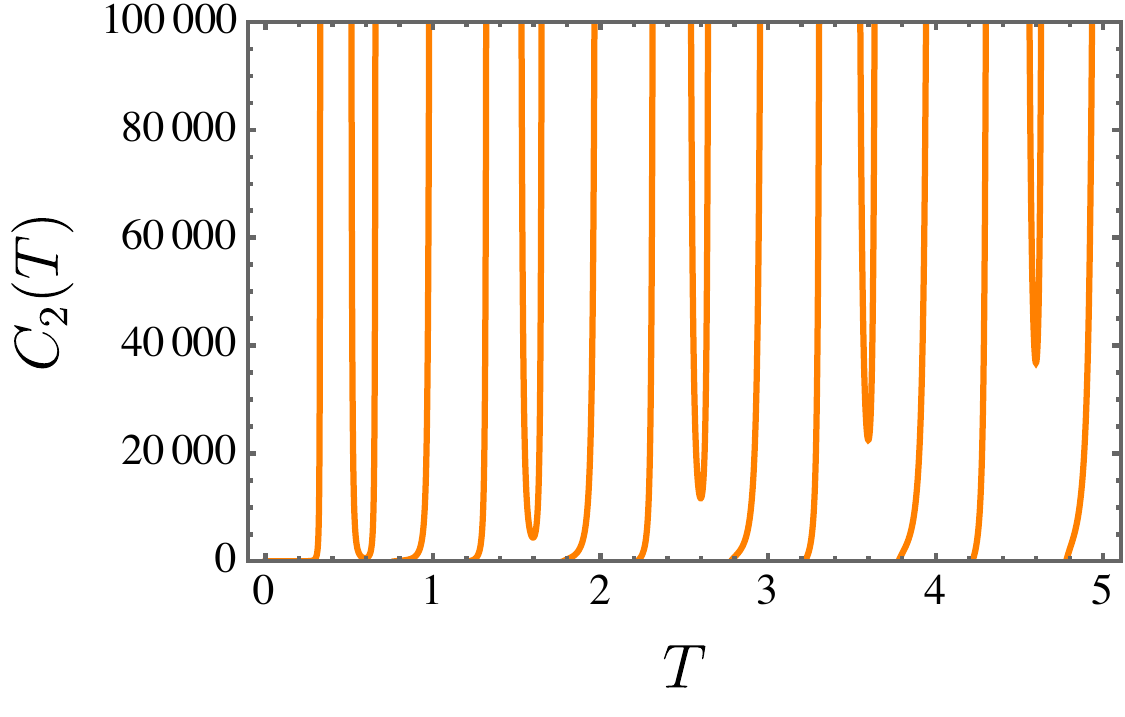}
    \includegraphics[scale=0.4]{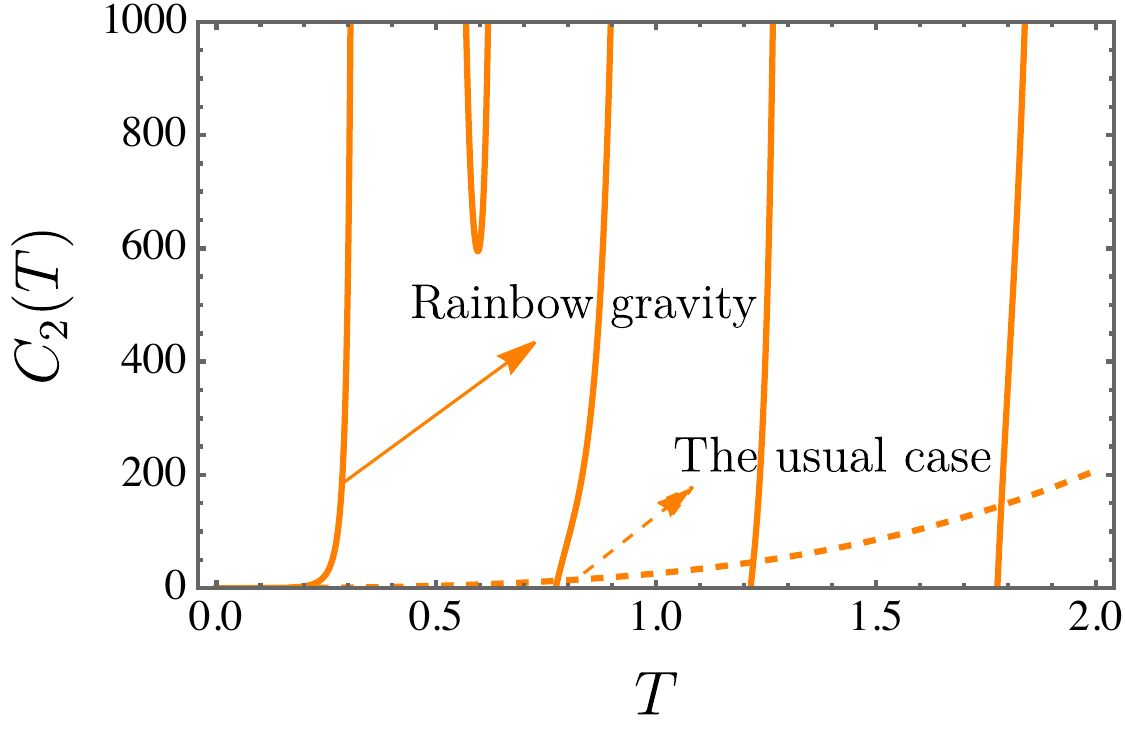}
    \caption{Heat capacity for different values of temperature $T$, considering the second case.}
    \label{HeatCapacity2}
\end{figure}


\subsubsection{Mean energy}

Understanding the behavior of particles at extremely high energy scales is a crucial area of research in physics, with important implications for our understanding of the universe. Studying the mean energy of an ensemble of particles in the context of \textit{rainbow} gravity can provide valuable insights into the distribution of energy among particles and the nature of spacetime at these scales. This can help us better understand the effects of quantum gravity and inform the development of new applications. 
Then, the mean energy reads,
\ie
\begin{split}
U_2(T) = &\int_{0}^{\infty } \frac{E_{P}^2 E \left(e^{\frac{E \xi }{E_{P}}}-1\right)^2 e^{\frac{E \xi }{E_{P}}-\beta  E}}{\xi ^2 \left(1-e^{-\beta  E}\right)} \, \mathrm{d}E \\
& = \frac{E_{P}^2 \left(\psi_{ 1}\left(1-\frac{\xi }{E_{P} \beta }\right)-2 \psi_{1}\left(1-\frac{2 \xi }{E_{P}\beta }\right)+\psi_{1}\left(1-\frac{3 \xi }{E_{P} \beta }\right)\right)}{\beta ^2 \xi ^2},
\end{split}
\fe
where $\psi_{n}(z)$ is the \textit{pollygamma function} defined by the derivative of the logarithm of the \textit{gamma function} $\Gamma(z) \equiv \int^{\infty}_{0} t^{z-1}e^{-t} \mathrm{d}t$, which can explicitly given by
\ie
\psi_{n}(z) = \frac{\mathrm{d}^{n+1}}{\mathrm{d}z^{n+1}}\ln[\Gamma(z)] = (-1)^{n+1} n! \sum^{\infty}_{k=0} \frac{1}{(z+k)^{n+1}} = (-1)^{n+1} n! \zeta(n+1,z), \,\,\,\, \forall n>0,
\fe
with $\zeta(a,z)$ being the \textit{Hurwitz zeta function}. Notice that, if we consider the limit where $\xi \rightarrow 0$, we obtain 
$\lim_{\xi \rightarrow 0}U_{2}(T) = \pi^{4}T^{4}/15$, recovering also the well--known \textit{Stefan--Boltzmann} law. Also, if the temperature $T \rightarrow \infty$, $\lim_{T \rightarrow 0}U_{2}(T) = 0$.

Special functions, have been used to analyze the thermodynamic properties of various systems, including black holes, and Bose--Einstein condensation (BEC) phase transition, and. For instance, in Refs. \cite{b1,b2,b3,b4}, the authors used the ensemble theory to analyze a trapped Bose gas, and calculated the thermodynamic potentials and critical temperature for BEC. In addition, they have also been used to study the the thermodynamical properties in different scenarios of gravitational theories, such as Gauss--Bonnet \cite{cvetivc2002black}, higher curvature gravities \cite{myers1988black,lov2,lov3}, holography \cite{dong2014holographic,hollogra}, and charged black holes \cite{zou2014critical}.

Very recently in the literature, $\psi_{n}(z)$ has also been employed to address an arbitrary number of dipoles at the sites of a regular one--dimensional crystal lattice \cite{ciftja2023exact}, geometric study of fluctuating $1/2$--BPS statistical configurations \cite{bellucci2010exact}, Tsales statistics \cite{niven2009q}, expansion of the one-loop corrections \cite{klajn2014exact}, correspondence between thermodynamics and inference \cite{lamont2019correspondence}, double wrapping in twisted AdS/CFT, Five loop Konishi from AdS/CFT \cite{bajnok2010five}, thermodynamics of Gaussian fluctuations and paraconductivity in layered superconductors\cite{mishonov2000thermodynamics}, Kurtosis of von Neumann entanglement entropy\cite{huang2021kurtosis}, and macroscopically ordered water in nanopores \cite{kofinger2008macroscopically}, and gas--liquid transition in the system of dipolar hard spheres \cite{levin1999happened}.

In Fig. \ref{Energy2}, the behavior the the mean energy might possibly indicate either quantum fluctuations of the massless particles or an extreme phase transition in this scenario.

\begin{figure}
    \centering
    \includegraphics[scale=0.4]{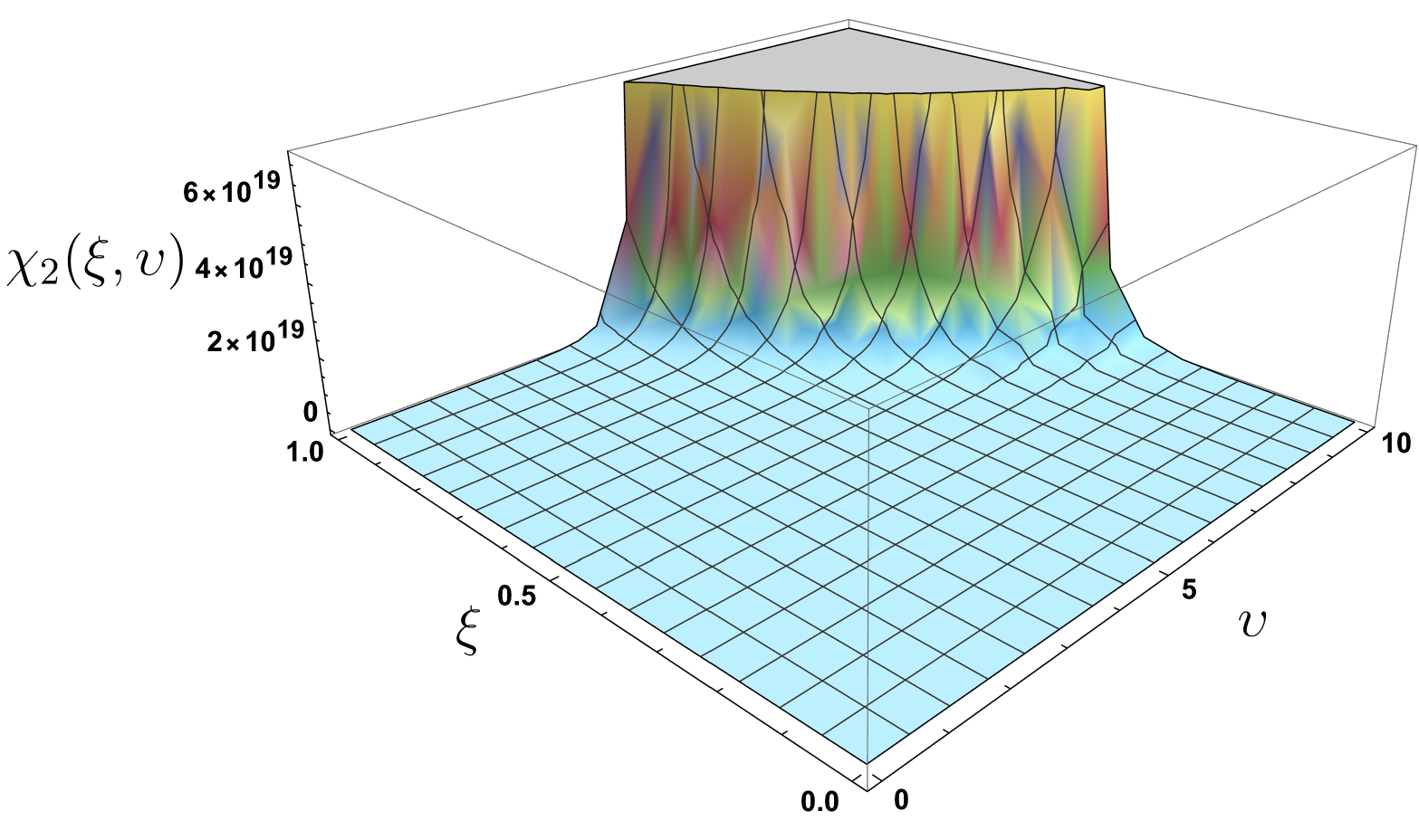}
    \caption{The black body radiation for different values of $\xi$ and $\upsilon$.}
    \label{radiation2}
\end{figure}

\subsubsection{Entropy}

The study of entropy in the context of \textit{rainbow} gravity is a crucial area of research in physics, as it provides a unique perspective on the behavior of particles in extreme conditions. By analyzing the distribution of entropy among particles, we can gain a deeper understanding of the fundamental nature of spacetime and its underlying quantum structure. This knowledge has important implications for the development of new theories of physics, as well as for practical applications in fields such as quantum computing and information theory. Moreover, studying the entropy of particle ensembles may provide novel perspectives on the behavior of matter and energy in high--energy regimes, which can ultimately lead to groundbreaking discoveries and advancements in our understanding of the universe. The entropy reads
\ie
\begin{split}
&S_{2}(T) =  \int_0^{\infty } \beta ^2 \left(\frac{E_{P}^2 E \left(e^{\frac{E \xi }{E_{P}}}-1\right)^2 e^{\frac{E \xi }{E_{P}}-\beta  E}}{\beta  \xi ^2 \left(1-e^{-\beta  E}\right)}-\frac{E_{P}^2 e^{\frac{E \xi }{E_{P}}} \left(e^{\frac{E \xi }{E_{P}}}-1\right)^2 \log \left(1-e^{-\beta  E}\right)}{\beta ^2 \xi ^2}\right) \, \mathrm{d}E\\
& = \frac{E_{P}^2}{36 \xi ^2} \left(-\frac{18 \beta ^2 E_{P}^3}{(E_{P} \beta -2 \xi ) \xi ^2}+\frac{4 \beta ^2 E_{P}^3}{(E_{P} \beta -3 \xi ) \xi ^2}-\frac{36 \beta ^2 E_{P}^3}{\xi  (\xi -E_{P} \beta )^2}+\frac{14 \beta  E_{P}^2}{\xi ^2} +\frac{54 \beta  E_{P}^2}{(\xi -E_{P} \beta )^2} \right.\\
& \left. -\frac{36 \beta  E_{P}^2}{(E_{P} \beta -2 \xi )^2}+\frac{18 \beta  E_{P}^2}{(E_{P} \beta -3 \xi )^2}-\frac{36 \gamma  \beta  E_{P}^2}{E_{P} \beta  \xi -\xi ^2}+\frac{72 \beta  E_{P}^2}{E_{P} \beta  \xi -2 \xi ^2}-\frac{24 \beta  E_{P}^2}{E_{P} \beta  \xi -3 \xi ^2} \right. \\
& \left. +\frac{36 \psi_{0}\left(2-\frac{2 \xi }{E_{P} \beta }\right)^2 E_{P}}{E_{P} \beta -2 \xi }+\frac{36 \psi_{0}\left(-\frac{\xi }{E_{P} \beta }\right) E_{P}}{E_{P} \beta -\xi }+36 \left(\frac{1}{\xi }-\frac{2}{E_{P} \beta -2 \xi }\right) \psi_{0}\left(-\frac{2 \xi }{E_{P} \beta }\right) E_{P} \right.\\
& \left. +\left(\frac{36}{E_{P} \beta -3 \xi }-\frac{12}{\xi }\right) \psi_{0}\left(-\frac{3 \xi }{E_{P} \beta }\right) E_{P}+\frac{18 \gamma ^2 E_{P}}{E_{P} \beta -\xi }+\frac{3 \pi ^2 E_{P}}{E_{P} \beta -\xi }-\frac{18 \psi_{0}\left(2-\frac{\xi }{E_{P} \beta }\right)^2 E_{P}}{E_{P} \beta -\xi }\right.\\
& \left. -\frac{36 \psi_{0}\left(-\frac{\xi }{E_{P} \beta }\right) E_{P}}{\xi }+\frac{24 \gamma  E_{P}}{\xi }  -\frac{36 \gamma ^2 E_{P}}{E_{P} \beta -2 \xi }-\frac{6 \pi ^2 E_{P}}{E_{P} \beta -2 \xi }+\frac{18 \gamma ^2 E_{P}}{E_{P} \beta -3 \xi }+\frac{3 \pi ^2 E_{P}}{E_{P} \beta -3 \xi } \right.\\
& \left. -\frac{18 \psi_{0}\left(2-\frac{3 \xi }{E_{P} \beta }\right)^2 E_{P}}{E_{P} \beta -3 \xi }  +\frac{18 \psi_{0}\left(1-\frac{\xi }{E_{P} \beta }\right)^2}{\beta }+\frac{18 \xi  \psi_{0}\left(2-\frac{\xi }{E_{P} \beta }\right)^2}{\beta  (E_{P} \beta -\xi )}+\frac{18 \psi_{0}\left(1-\frac{3 \xi }{E_{P} \beta }\right)^2}{\beta }  \right. \\
& \left. +\frac{54 \xi  \psi_{0}\left(2-\frac{3 \xi }{E_{P} \beta }\right)^2}{\beta  (E_{P} \beta -3 \xi )} + \frac{36 \psi_{1}\left(-\frac{\xi }{E_{P} \beta }\right)}{\beta }+\frac{36 \psi_{1}\left(-\frac{3 \xi }{E_{P} \beta }\right)}{\beta }-\frac{36 \psi_{0}\left(1-\frac{2 \xi }{E_{P} \beta }\right)^2}{\beta }\right.\\
& \left. -\frac{72 \psi_{1}\left(-\frac{2 \xi }{E_{P} \beta }\right)}{\beta }-\frac{72 \xi  \psi_{0}\left(2-\frac{2 \xi }{E_{P} \beta }\right)^2}{\beta  (E_{P} \beta -2 \xi )} -\frac{18 \gamma ^2 \xi }{E_{P} \beta ^2-\beta  \xi }-\frac{3 \pi ^2 \xi }{E_{P} \beta ^2-\beta  \xi }+\frac{72 \gamma ^2 \xi }{E_{P} \beta ^2-2 \beta  \xi } \right. \\
& \left. +\frac{12 \pi ^2 \xi }{E_{P} \beta ^2-2 \beta  \xi }-\frac{54 \gamma ^2 \xi }{E_{P} \beta ^2-3 \beta  \xi }-\frac{9 \pi ^2 \xi }{E_{P} \beta ^2-3 \beta  \xi } +\frac{36 \gamma  E_{P}}{E_{P} \beta -\xi } \right),
\end{split}
\fe
where, $\gamma$ is the Euler gamma defined by $\gamma = \lim_{n\rightarrow \infty} \left(\sum^{n}_{k=1} \frac{1}{k} - \ln(n) \right)$. Here, if we consider the limit where $\lim_{\xi \rightarrow 0}S_{2}(T) = T^{3}[4 \pi ^4-45 \gamma  (2+\psi ^{(2)}(1)-\psi ^{(2)}(2))+45 (2+\psi ^{(2)}(1)-\psi ^{(2)}(2))/45] = 4\pi^{4}T^{3}/45$. Also, for $\lim_{T \rightarrow 0}S_{2}(T) = 0$. The behavior of the entropy for this case is displayed in Fig. \ref{Entropy2}, comparing such thermal function with the usal case.

\subsubsection{Heat capacity}

Finally, in this subsection we provide the study of the heat capacity. In this way, we write
\ie
\begin{split}
C_{2}(T)& =\int_{0}^{\infty }-\beta ^{2}\left( -\frac{E_{P}^{2}E^{2}\left(
e^{\frac{E\xi }{E_{P}}}-1\right) ^{2}e^{\frac{E\xi }{E_{P}}-\beta E}}{\xi
^{2}\left( 1-e^{-\beta E}\right) }-\frac{E_{P}^{2}E^{2}\left( e^{\frac{E\xi 
}{E_{P}}}-1\right) ^{2}e^{\frac{E\xi }{E_{P}}-2\beta E}}{\xi ^{2}\left(
1-e^{-\beta E}\right) ^{2}}\right) \,\mathrm{d}E \\
& =\frac{E_{P}}{\beta ^{2}\xi ^{2}}\left\{ 2E_{P}\beta \left[ \psi
_{1}\left( -\frac{\xi }{E_{P}\beta }\right) -2\psi _{1}\left( -\frac{2\xi }{%
E_{P}\beta }\right) +\psi _{1}\left( -\frac{3\xi }{E_{P}\beta }\right) %
\right] \right.  \\
& \left. -\xi \left[ \psi _{2}\left( -\frac{\xi }{E_{P}\beta }\right) -4\psi
_{2}\left( -\frac{2\xi }{E_{P}\beta }\right) +3\psi _{2}\left( -\frac{3\xi }{%
E_{P}\beta }\right) \right] \right\} 
\end{split}%
\fe
In addition, we have $\lim_{\rightarrow 0}C_{2}(T) = 4\pi^{4}T^{3}/15$. Also, for $\lim_{T \rightarrow 0}C_{2}(T) = 0$. Finally, we compare the role of \textit{rainbow} parameters in comparison with the usual case. This analysis is shown in  Fig. \ref{HeatCapacity2}.



\section{Conclusion} \label{conclusion}

This study was devoted to studying the thermodynamic behavior of photon-like particles within the framework of \textit{rainbow} gravity. To accomplish this, two particular ansatzs were chosen for conducting the calculations. Initially, a dispersion relation was considered that avoided UV divergences, resulting in a positive effective cosmological constant. \textit{Numerical} analysis was performed to evaluate the thermodynamic functions of the system and estimate bounds. Furthermore, a phase transition was also observed in this model.

Next, a dispersion relation employed in the context of \textit{Gamma Ray Bursts} was investigated. Interestingly, in this case, the thermodynamic properties were calculated \textit{analytically}. It was found that these properties depended on various mathematical functions, such as the harmonic series $H_{n}$, gamma $\Gamma(z)$, polygamma $\psi_{n}(z)$ and zeta Riemann functions $\zeta(z)$. Overall, this work contributed to the understanding of the thermodynamics of photon--like particles within the \textit{rainbow} gravity formalism. The combination of \textit{numerical} and \textit{analytical} approaches provided valuable insights into the behavior of the system and established connections with important mathematical functions. Further exploration of these phenomena could deepen our comprehension of fundamental physics and its implications.

As a future perspective, analyzing the role of the mass within this context seems to be a interesting question worthy to be investigated.


\section*{Acknowledgments}
\hspace{0.5cm}

The authors thank CNPq and CAPES (Brazilian research agencies) for their financial support. Particularly, A. A. Araújo Filho is supported by Conselho Nacional de Desenvolvimento Cientíıfico e Tecnológico (CNPq) -- 200486/2022-5. JF would like to thank the Fundação Cearense de Apoio ao Desenvolvimento Cient\'{i}fico e Tecnol\'{o}gico (FUNCAP) under the grant PRONEM PNE0112-00085.01.00/16 for financial support.


\section{Data Availability Statement}

Data Availability Statement: No Data associated in the manuscript


\bibliographystyle{ieeetr}
\bibliography{main}

\begin{thebibliography}{10}

\bibitem{majumder2013singularity}
B.~Majumder, ``Singularity free rainbow universe,'' {\em International Journal
  of Modern Physics D}, vol.~22, no.~12, p.~1342021, 2013.

\bibitem{kangal2022effective}
E.~Kangal, K.~Sogut, M.~Salti, and O.~Aydogdu, ``Effective dynamics of spin-1/2
  particles in a rainbow universe,'' {\em Annals of Physics}, vol.~444,
  p.~169018, 2022.

\bibitem{kangal2021relativistic}
E.~Kangal, M.~Salti, O.~Aydogdu, and K.~Sogut, ``Relativistic quantum dynamics
  of scalar particles in the rainbow formalism of gravity,'' {\em Physica
  Scripta}, vol.~96, no.~9, p.~095301, 2021.

\bibitem{r1}
A.~F. Ali and M.~M. Khalil, ``A proposal for testing gravity's rainbow,'' {\em
  Europhysics Letters}, vol.~110, no.~2, p.~20009, 2015.

\bibitem{r2}
S.~Gangopadhyay and A.~Dutta, ``Constraints on rainbow gravity functions from
  black-hole thermodynamics,'' {\em Europhysics Letters}, vol.~115, no.~5,
  p.~50005, 2016.

\bibitem{r3}
A.~Awad, A.~F. Ali, and B.~Majumder, ``Nonsingular rainbow universes,'' {\em
  Journal of Cosmology and Astroparticle Physics}, vol.~2013, no.~10, p.~052,
  2013.

\bibitem{r4}
A.~F. Ali, M.~Faizal, B.~Majumder, and R.~Mistry, ``Gravitational collapse in
  gravity's rainbow,'' {\em International Journal of Geometric Methods in
  Modern Physics}, vol.~12, no.~09, p.~1550085, 2015.

\bibitem{r5}
M.~de~Montigny, J.~Pinfold, S.~Zare, and H.~Hassanabadi, ``Klein--gordon
  oscillator in a global monopole space--time with rainbow gravity,'' {\em The
  European Physical Journal Plus}, vol.~137, pp.~1--17, 2022.

\bibitem{r6}
J.~Magueijo, ``New varying speed of light theories,'' {\em Reports on Progress
  in Physics}, vol.~66, no.~11, p.~2025, 2003.

\bibitem{r8}
G.~F. Ellis, ``Note on varying speed of light cosmologies,'' {\em arXiv
  preprint astro-ph/0703751}, 2007.

\bibitem{r9}
G.~F. Ellis and J.-P. Uzan, ``c is the speed of light, isn’t it?,'' {\em
  American journal of physics}, vol.~73, no.~3, pp.~240--247, 2005.

\bibitem{sefiedgar2017entropic}
A.~Sefiedgar, ``From the entropic force to the friedmann equation in rainbow
  gravity,'' {\em Europhysics Letters}, vol.~117, no.~6, p.~69001, 2017.

\bibitem{brighenti2017primordial}
F.~Brighenti, G.~Gubitosi, and J.~Magueijo, ``Primordial perturbations in a
  rainbow universe with running newton constant,'' {\em Physical Review D},
  vol.~95, no.~6, p.~063534, 2017.

\bibitem{nojiri2007introduction}
S.~Nojiri and S.~D. Odintsov, ``Introduction to modified gravity and
  gravitational alternative for dark energy,'' {\em International Journal of
  Geometric Methods in Modern Physics}, vol.~4, no.~01, pp.~115--145, 2007.

\bibitem{amelino2013rainbow}
G.~Amelino-Camelia, M.~Arzano, G.~Gubitosi, and J.~Magueijo, ``Rainbow gravity
  and scale-invariant fluctuations,'' {\em Physical Review D}, vol.~88, no.~4,
  p.~041303, 2013.

\bibitem{sefiedgar2018thermodynamics}
A.~Sefiedgar and M.~Mirzazadeh, ``Thermodynamics of the frw universe at the
  event horizon in palatini f(r) gravity,'' {\em Advances in High Energy
  Physics}, vol.~2018, pp.~1--6, 2018.

\bibitem{mota2019combined}
C.~E. Mota, L.~C. Santos, G.~Grams, F.~M. da~Silva, and D.~P. Menezes,
  ``Combined rastall and rainbow theories of gravity with applications to
  neutron stars,'' {\em Physical Review D}, vol.~100, no.~2, p.~024043, 2019.

\bibitem{younesizadeh2021new}
Y.~Younesizadeh, A.~H. Ahmed, A.~A. Ahmad, Y.~Younesizadeh, and M.~Ebrahimkhas,
  ``New class of solutions in $f (r)$-gravity's rainbow and $f(r)$-gravity:
  Exact solutions thermodynamics quasinormal modes,'' {\em Nuclear Physics B},
  vol.~971, p.~115376, 2021.

\bibitem{fomin2020exact}
I.~Fomin and S.~Chervon, ``Exact and slow-roll solutions for exponential
  power-law inflation connected with modified gravity and observational
  constraints,'' {\em Universe}, vol.~6, no.~11, p.~199, 2020.

\bibitem{dehghani2018thermal}
M.~Dehghani, ``Thermal fluctuations of dilaton black holes in gravity's
  rainbow,'' {\em Physics Letters B}, vol.~781, pp.~553--560, 2018.

\bibitem{dehghani2019thermodynamic}
M.~Dehghani, ``Thermodynamic properties of novel dilatonic btz black holes
  under the influence of rainbow gravity,'' {\em Physics Letters B}, vol.~799,
  p.~135037, 2019.

\bibitem{sefiedgar2017thermodynamics}
A.~S. Sefiedgar and M.~Daghigh, ``Thermodynamics of the frw universe in rainbow
  gravity,'' {\em International Journal of Modern Physics D}, vol.~26, no.~13,
  p.~1750139, 2017.

\bibitem{haldar2019thermodynamics}
A.~Haldar and R.~Biswas, ``Thermodynamics of reissner--nordstr o{\"{}} m black
  holes in higher dimensions: rainbow gravity background with general
  uncertainty principle,'' {\em General Relativity and Gravitation}, vol.~51,
  no.~6, p.~72, 2019.

\bibitem{hamil2022effect}
B.~Hamil and B.~L{\"u}tf{\"u}o{\u{g}}lu, ``Effect of snyder--de sitter model on
  the black hole thermodynamics in the context of rainbow gravity,'' {\em
  International Journal of Geometric Methods in Modern Physics}, vol.~19,
  no.~03, p.~2250047, 2022.

\bibitem{kim2016thermodynamic}
Y.-W. Kim, S.~K. Kim, and Y.-J. Park, ``Thermodynamic stability of modified
  schwarzschild--ads black hole in rainbow gravity,'' {\em The European
  Physical Journal C}, vol.~76, pp.~1--11, 2016.

\bibitem{dehghani2018thermodynamics}
M.~Dehghani, ``Thermodynamics of charged dilatonic btz black holes in rainbow
  gravity,'' {\em Physics Letters B}, vol.~777, pp.~351--360, 2018.

\bibitem{araujo2021bouncing}
A.~A. Ara{\'u}jo~Filho and A.~Y. Petrov, ``Bouncing universe in a heat bath,''
  {\em International Journal of Modern Physics A}, vol.~36, no.~34n35,
  p.~2150242, 2021.

\bibitem{feng2017thermodynamic}
Z.-W. Feng and S.-Z. Yang, ``Thermodynamic phase transition of a black hole in
  rainbow gravity,'' {\em Physics Letters B}, vol.~772, pp.~737--742, 2017.

\bibitem{dehghani2020ads4}
M.~Dehghani, ``Ads4 black holes with nonlinear source in rainbow gravity,''
  {\em Physics Letters B}, vol.~801, p.~135191, 2020.

\bibitem{md2018phase}
S.~Md, ``Phase transition of quantum-corrected schwarzschild black hole in
  rainbow gravity,'' {\em Physics Letters B}, vol.~784, pp.~6--11, 2018.

\bibitem{Furtado:2021aod}
J.~Furtado, J.~F. Assun\c{c}\~ao, and C.~R. Muniz, ``{Relativistic
  Bose-Einstein condensate in the rainbow gravity},'' {\em EPL}, vol.~139,
  p.~29001, 2022.

\bibitem{nilsson2017energy}
N.~A. Nilsson and M.~P. Dabrowski, ``Energy scale of lorentz violation in
  rainbow gravity,'' {\em Physics of the Dark Universe}, vol.~18, pp.~115--122,
  2017.

\bibitem{roy2023entropy}
T.~Roy and U.~Debnath, ``Entropy bound and egup correction of d-dimensional
  reissner-nordstrom black hole in rainbow gravity,'' {\em International
  Journal of Modern Physics A}, 2023.

\bibitem{feng2020rainbow}
Z.-W. Feng, X.~Zhou, S.-Q. Zhou, and D.-D. Feng, ``Rainbow gravity corrections
  to the information flux of a black hole and the sparsity of hawking
  radiation,'' {\em Annals of Physics}, vol.~416, p.~168144, 2020.

\bibitem{yekta2019joule}
D.~M. Yekta, A.~Hadikhani, and {\"O}.~{\"O}kc{\"u}, ``Joule-thomson expansion
  of charged ads black holes in rainbow gravity,'' {\em Physics Letters B},
  vol.~795, pp.~521--527, 2019.

\bibitem{feng2018rainbow}
Z.-W. Feng and S.-Z. Yang, ``Rainbow gravity corrections to the entropic
  force,'' {\em Advances in High Energy Physics}, vol.~2018, 2018.

\bibitem{sefiedgar2016can}
A.~S. Sefiedgar, ``How can rainbow gravity affect gravitational force?,'' {\em
  International Journal of Modern Physics D}, vol.~25, no.~14, p.~1650101,
  2016.

\bibitem{amelino2001testable}
G.~Amelino-Camelia, ``Testable scenario for relativity with minimum length,''
  {\em Physics Letters B}, vol.~510, no.~1-4, pp.~255--263, 2001.

\bibitem{amelino2002relativity}
G.~Amelino-Camelia, ``Relativity in spacetimes with short-distance structure
  governed by an observer-independent (planckian) length scale,'' {\em
  International Journal of Modern Physics D}, vol.~11, no.~01, pp.~35--59,
  2002.

\bibitem{magueijo2004gravity}
J.~Magueijo and L.~Smolin, ``Gravity's rainbow,'' {\em Classical and Quantum
  Gravity}, vol.~21, no.~7, p.~1725, 2004.

\bibitem{garattini2010modified}
R.~Garattini, ``Modified dispersion relations and black hole entropy,'' {\em
  Physics Letters B}, vol.~685, no.~4-5, pp.~329--337, 2010.

\bibitem{garattini2011modified}
R.~Garattini and G.~Mandanici, ``Modified dispersion relations lead to a finite
  zero point gravitational energy,'' {\em Physical Review D}, vol.~83, no.~8,
  p.~084021, 2011.

\bibitem{garattini2005casimir}
R.~Garattini, ``Casimir energy, the cosmological constant and massive
  gravitons,'' {\em arXiv preprint gr-qc/0510062}, 2005.

\bibitem{garattini2006cosmological}
R.~Garattini, ``The cosmological constant as an eigenvalue of a
  sturm--liouville problem and its renormalization,'' {\em Journal of Physics
  A: Mathematical and General}, vol.~39, no.~21, p.~6393, 2006.

\bibitem{BBRadiations}
K.~Konar, K.~Bose, and R.~Paul, ``Revisiting cosmic microwave background
  radiation using blackbody radiation inversion,'' {\em Scientific Reports},
  vol.~11, no.~1, p.~1008, 2021.

\bibitem{araujo2023thermodynamics}
A.~A. Ara{\'u}jo~Filho, S.~Zare, P.~Porf{\'\i}rio, J.~K{\v{r}}{\'\i}{\v{z}},
  and H.~Hassanabadi, ``Thermodynamics and evaporation of a modified
  schwarzschild black hole in a non--commutative gauge theory,'' {\em Physics
  Letters B}, p.~137744, 2023.

\bibitem{furtado2023thermodynamical}
J.~Furtado, J.~Silva, {\em et~al.}, ``Thermodynamical properties of an ideal
  gas in a traversable wormhole,'' {\em arXiv preprint arXiv:2302.05492}, 2023.

\bibitem{aa1}
R.~R. Oliveira, A.~A. Ara{\'u}jo~Filho, F.~C. Lima, R.~V. Maluf, and C.~A.
  Almeida, ``Thermodynamic properties of an aharonov-bohm quantum ring,'' {\em
  The European Physical Journal Plus}, vol.~134, no.~10, p.~495, 2019.

\bibitem{aa2}
A.~A. Ara{\'u}jo~Filho and J.~Reis, ``Thermal aspects of interacting quantum
  gases in lorentz-violating scenarios,'' {\em The European Physical Journal
  Plus}, vol.~136, pp.~1--30, 2021.

\bibitem{aa3}
R.~Oliveira {\em et~al.}, ``Thermodynamic properties of neutral dirac particles
  in the presence of an electromagnetic field,'' {\em The European Physical
  Journal Plus}, vol.~135, no.~1, pp.~1--10, 2020.

\bibitem{aa4}
A.~A. Ara{\'u}jo~Filho, ``Lorentz-violating scenarios in a thermal reservoir,''
  {\em The European Physical Journal Plus}, vol.~136, no.~4, pp.~1--14, 2021.

\bibitem{aa5}
R.~Oliveira, A.~A. Ara{\'u}jo~Filho, R.~Maluf, and C.~Almeida, ``The
  relativistic aharonov--bohm--coulomb system with position-dependent mass,''
  {\em Journal of Physics A: Mathematical and Theoretical}, vol.~53, no.~4,
  p.~045304, 2020.

\bibitem{aa6}
A.~A. Ara{\'u}jo~Filho and R.~V. Maluf, ``Thermodynamic properties in
  higher-derivative electrodynamics,'' {\em Brazilian Journal of Physics},
  vol.~51, pp.~820--830, 2021.

\bibitem{aa7}
A.~A. Ara{\'u}jo~Filho and A.~Y. Petrov, ``Higher-derivative lorentz-breaking
  dispersion relations: a thermal description,'' {\em The European Physical
  Journal C}, vol.~81, no.~9, p.~843, 2021.

\bibitem{aa8}
A.~A. Ara{\'u}jo~Filho and A.~Y. Petrov, ``Bouncing universe in a heat bath,''
  {\em International Journal of Modern Physics A}, vol.~36, no.~34n35,
  p.~2150242, 2021.

\bibitem{aa9}
A.~A. Ara{\'u}jo~Filho, ``Thermodynamics of massless particles in curved
  spacetime,'' {\em arXiv preprint arXiv:2201.00066}, 2022.

\bibitem{aa10}
A.~A. Ara{\'u}jo~Filho, ``Particles in loop quantum gravity formalism: a
  thermodynamical description,'' {\em Annalen der Physik}, p.~2200383, 2022.

\bibitem{aa11}
A.~A. Ara{\'u}jo~Filho, J.~Reis, and S.~Ghosh, ``Fermions on a torus knot,''
  {\em The European Physical Journal Plus}, vol.~137, no.~5, p.~614, 2022.

\bibitem{aa12}
A.~A. Ara{\'u}jo~Filho and J.~Reis, ``How does geometry affect quantum
  gases?,'' {\em International Journal of Modern Physics A}, vol.~37,
  no.~11n12, p.~2250071, 2022.

\bibitem{aa13}
P.~Sedaghatnia, H.~Hassanabadi, J.~Porf{\'\i}rio, W.~Chung, {\em et~al.},
  ``Thermodynamical properties of a deformed schwarzschild black hole via dunkl
  generalization,'' {\em arXiv preprint arXiv:2302.11460}, 2023.

\bibitem{aa14}
A.~A. Ara{\'u}jo~Filho, J.~Furtado, and J.~Silva, ``Thermodynamical properties
  of an ideal gas in a traversable wormhole,'' {\em arXiv preprint
  arXiv:2302.05492}, 2023.

\bibitem{51}
G.~Amelino-Camelia, J.~Ellis, N.~Mavromatos, D.~V. Nanopoulos, and S.~Sarkar,
  ``Tests of quantum gravity from observations of $\gamma$-ray bursts,'' {\em
  Nature}, vol.~393, no.~6687, pp.~763--765, 1998.

\bibitem{52}
A.~Awad, A.~F. Ali, and B.~Majumder, ``Nonsingular rainbow universes,'' {\em
  Journal of Cosmology and Astroparticle Physics}, vol.~2013, no.~10, p.~052,
  2013.

\bibitem{53}
G.~Santos, G.~Gubitosi, and G.~Amelino-Camelia, ``On the initial singularity
  problem in rainbow cosmology,'' {\em Journal of Cosmology and Astroparticle
  Physics}, vol.~2015, no.~08, p.~005, 2015.

\bibitem{b1}
G.~Baym and C.~J. Pethick, ``Ground-state properties of magnetically trapped
  bose-condensed rubidium gas,'' {\em Physical review letters}, vol.~76, no.~1,
  p.~6, 1996.

\bibitem{b2}
E.~H. Lieb and R.~Seiringer, ``Proof of bose-einstein condensation for dilute
  trapped gases,'' {\em Physical review letters}, vol.~88, no.~17, p.~170409,
  2002.

\bibitem{b3}
L.~Santos, G.~Shlyapnikov, P.~Zoller, and M.~Lewenstein, ``Bose-einstein
  condensation in trapped dipolar gases,'' {\em Physical Review Letters},
  vol.~85, no.~9, p.~1791, 2000.

\bibitem{b4}
F.~Dalfovo, S.~Giorgini, L.~P. Pitaevskii, and S.~Stringari, ``Theory of
  bose-einstein condensation in trapped gases,'' {\em Reviews of modern
  physics}, vol.~71, no.~3, p.~463, 1999.

\bibitem{cvetivc2002black}
M.~Cveti{\v{c}}, S.~Nojiri, and S.~D. Odintsov, ``Black hole thermodynamics and
  negative entropy in de sitter and anti-de sitter einstein--gauss--bonnet
  gravity,'' {\em Nuclear Physics B}, vol.~628, no.~1-2, pp.~295--330, 2002.

\bibitem{myers1988black}
R.~C. Myers and J.~Z. Simon, ``Black-hole thermodynamics in lovelock gravity,''
  {\em Physical Review D}, vol.~38, no.~8, p.~2434, 1988.

\bibitem{lov2}
H.~Saida and J.~Soda, ``Statistical entropy of btz black hole in higher
  curvature gravity,'' {\em Physics Letters B}, vol.~471, no.~4, pp.~358--366,
  2000.

\bibitem{lov3}
S.~Banerjee, A.~Bhattacharyya, A.~Kaviraj, K.~Sen, and A.~Sinha, ``Constraining
  gravity using entanglement in ads/cft,'' {\em Journal of High Energy
  Physics}, vol.~2014, no.~5, pp.~1--34, 2014.

\bibitem{dong2014holographic}
X.~Dong, ``Holographic entanglement entropy for general higher derivative
  gravity,'' {\em Journal of High Energy Physics}, vol.~2014, no.~1, pp.~1--32,
  2014.

\bibitem{hollogra}
R.-X. Miao, ``An exact construction of codimension two holography,'' {\em
  Journal of High Energy Physics}, vol.~2021, no.~1, pp.~1--27, 2021.

\bibitem{zou2014critical}
D.-C. Zou, Y.~Liu, and B.~Wang, ``Critical behavior of charged gauss-bonnet-ads
  black holes in the grand canonical ensemble,'' {\em Physical Review D},
  vol.~90, no.~4, p.~044063, 2014.

\bibitem{ciftja2023exact}
O.~Ciftja, ``Exact ground state energy of a system with an arbitrary number of
  dipoles at the sites of a regular one-dimensional crystal lattice,'' {\em
  Journal of Physics and Chemistry of Solids}, vol.~172, p.~111044, 2023.

\bibitem{bellucci2010exact}
S.~Bellucci and B.~Nath~Tiwari, ``An exact fluctuating 1/2-bps configuration,''
  {\em Journal of High Energy Physics}, vol.~2010, no.~5, pp.~1--35, 2010.

\bibitem{niven2009q}
R.~K. Niven and H.~Suyari, ``The q-gamma and (q, q)-polygamma functions of
  tsallis statistics,'' {\em Physica A: Statistical Mechanics and its
  Applications}, vol.~388, no.~19, pp.~4045--4060, 2009.

\bibitem{klajn2014exact}
B.~Klajn, ``Exact high temperature expansion of the one-loop thermodynamic
  potential with complex chemical potential,'' {\em Physical Review D},
  vol.~89, no.~3, p.~036001, 2014.

\bibitem{lamont2019correspondence}
C.~H. LaMont and P.~A. Wiggins, ``Correspondence between thermodynamics and
  inference,'' {\em Physical Review E}, vol.~99, no.~5, p.~052140, 2019.

\bibitem{bajnok2010five}
Z.~Bajnok, {\'A}.~Heged{\H{u}}s, R.~A. Janik, and T.~{\L}ukowski, ``Five loop
  konishi from ads/cft,'' {\em Nuclear physics B}, vol.~827, no.~3,
  pp.~426--456, 2010.

\bibitem{mishonov2000thermodynamics}
T.~Mishonov and E.~Penev, ``Thermodynamics of gaussian fluctuations and
  paraconductivity in layered superconductors,'' {\em International Journal of
  Modern Physics B}, vol.~14, no.~32, pp.~3831--3879, 2000.

\bibitem{huang2021kurtosis}
Y.~Huang, L.~Wei, and B.~Collaku, ``Kurtosis of von neumann entanglement
  entropy,'' {\em Journal of Physics A: Mathematical and Theoretical}, vol.~54,
  no.~50, p.~504003, 2021.

\bibitem{kofinger2008macroscopically}
J.~K{\"o}finger, G.~Hummer, and C.~Dellago, ``Macroscopically ordered water in
  nanopores,'' {\em Proceedings of the National Academy of Sciences}, vol.~105,
  no.~36, pp.~13218--13222, 2008.

\bibitem{levin1999happened}
Y.~Levin, ``What happened to the gas-liquid transition in the system of dipolar
  hard spheres?,'' {\em Physical review letters}, vol.~83, no.~6, p.~1159,
  1999.

\end{thebibliography}

\end{document}